\documentclass[aps,preprint,groupaddress,showpacs]{revtex4}
 
\catcode`@=11  
\newcount\@tempcntc  
\def\@citex[#1]#2{\if@filesw\immediate\write\@auxout{\string\citation{#2}}\fi  
  \@tempcnta\z@\@tempcntb\m@ne\def\@citea{}\@cite{\@for\@citeb:=#2\do  
    {\@ifundefined  
       {b@\@citeb}{\@citeo\@tempcntb\m@ne\@citea\def\@citea{,}{\bf ?}\@warning  
       {Citation `\@citeb' on page \thepage \space undefined}}%
    {\setbox\z@\hbox{\global\@tempcntc0\csname b@\@citeb\endcsname\relax}%
     \ifnum\@tempcntc=\z@ \@citeo\@tempcntb\m@ne  
       \@citea\def\@citea{,}\hbox{\csname b@\@citeb\endcsname}%
     \else  
      \advance\@tempcntb\@ne  
      \ifnum\@tempcntb=\@tempcntc  
      \else\advance\@tempcntb\m@ne\@citeo  
      \@tempcnta\@tempcntc\@tempcntb\@tempcntc\fi\fi}}\@citeo}{#1}}  
\def\@citeo{\ifnum\@tempcnta>\@tempcntb\else\@citea\def\@citea{,}%
  \ifnum\@tempcnta=\@tempcntb\the\@tempcnta\else  
   {\advance\@tempcnta\@ne\ifnum\@tempcnta=\@tempcntb \else \def\@citea{--}\fi  
    \advance\@tempcnta\m@ne\the\@tempcnta\@citea\the\@tempcntb}\fi\fi}  
\catcode`@=12  
  
\begin{document}  
  
\preprint{DESY~11--014\hspace{12cm}ISSN 0418--9833} 
\preprint{LTH~904\hspace{15.6cm}}
 
\boldmath 
\title{Quark mixing renormalization effects in the determination of
$|V_{tq}|$}  
\unboldmath
  
\author{Andrea A. Almasy}
\email{a.a.almasy@liverpool.ac.uk} 
\affiliation{Department of Mathematical Sciences, University of Liverpool, Liverpool L69 3BX, UK}
\author{Bernd A. Kniehl}
\email{kniehl@desy.de}
\affiliation{{II.} Institut f\"ur Theoretische Physik,
Universit\"at Hamburg, Luruper Chaussee 149, 22761 Hamburg, Germany}
\author{Alberto Sirlin}
\email{alberto.sirlin@nyu.edu}
\affiliation{Department of Physics, New York University,
4 Washington Place, New York, New York 10003, USA}

\date{\today}

\begin{abstract}  
We study the numerical effects of several renormalization schemes of the Cabibbo-Kobayashi-Maskawa (CKM) quark mixing matrix on the top-quark decay widths. We then employ these results to infer the relative shifts in the CKM parameters $|V_{tq}|^2$ due to the quark mixing renormalization corrections, assuming that they are determined directly from the top-quark partial decay widths, without imposing unitarity constraints. We also discuss the implications of these effects on the ratio ${\cal R}=\Gamma(t\rightarrow Wb)/\Gamma_t$ and the determination of $|V_{tb}|^2$.
\end{abstract}  

\pacs{11.10.Gh,12.15.Hh,12.15.Lk,14.65.Ha}
  
\maketitle    

\section{Introduction}

The top quark is unique among the known quarks in that it decays before it can form hadronic bound states. This has important consequences. Above all, it offers us the possibility to explore the interactions of an unconfined quark at energies of a few hundred GeV to several TeV. Furthermore, it is an important asset of top-quark physics that the effects of both the electroweak  and strong interactions can, in most situations, be reliably evaluated. Needless to say, this is necessary for the analysis and interpretation of present and future experimental data.

The results from the Fermilab Tevatron are in accord with expectations and predictions of the Standard Model (SM). While the mass of this particle has been precisely measured, other properties and its production and decay dynamics have not been investigated in great detail so far. Hopefully this will change in the future. There are exciting physics topics to be explored. We mention here only a few of them. In view of its large mass, the top quark is an excellent probe of the mechanism that breaks the electroweak gauge symmetry and should, therefore, play a key role in clarifying the nature of the force(s) and particle(s) responsible for this phenomenon. The top quark is also a good probe for possible new parity-violating and/or non-SM CP-violating interactions, which could be induced, for instance, by non-standard Higgs bosons. Are there new top-quark decay modes, for instance, to supersymmetric particles? So far, experimental data are consistent with the SM prediction that $t\rightarrow Wb$ is the dominant decay mode, but its branching ratio and the structure of the $Wtb$ vertex are not yet measured directly with high accuracy. On the other hand, within the SM, the experimental measurement of the decay rate $\Gamma(t\rightarrow Wb)$ gives a direct determination of the $V_{tb}$ element of the Cabibbo-Kobayashi-Maskawa (CKM) quark mixing matrix. These and other topics, addressed at the Tevatron, will also be an important objective of future experiments at the CERN Large Hadron Collider (LHC).

Presently, the direct observation of the top quark at the Tevatron implies that $V_{tb}$ is known with a large, $15\%$ error~\cite{Aco2005,Aal2010,Aba2006,Aba2008,Aba2009}. The Particle Data Group~\cite{PDG2010} gives $V_{tb}$ with a much smaller error, using however the CKM unitarity constraints. At the LHC, one expects to extract $V_{tb}$ by direct observation with an error of ${\cal O}(5\%)$~\cite{AFGGHKKLM2007}, without appealing to CKM unitarity. 

In order to provide a precise theoretical framework for the direct determination of $|V_{tq}|$ ($q=d,s,b$), it is important to calculate accurately the top-quark partial decay widths and to analyze the phenomenological implications of the CKM matrix renormalization. The purpose of this paper is to carry out such an analysis, which complements our previous study of hadronic $W$-boson decays~\cite{AKS2008}.

This paper is organized as follows. In Sec.~\ref{top-decay}, we study numerically the effects of CKM matrix renormalization on the partial decay widths of the top quark at one loop, using the CKM matrix elements $V_{ij}$ obtained in the global analysis of the SM~\cite{PDG2010}. Similar to the study of the $W$-boson decay widths~\cite{AKS2008}, we focus on the prescriptions of Refs.~\cite{DS1990,GGM1999,DK2001,KS2006,KS2009}, which we also compare to the $\overline{\rm MS}$ scheme. These results are applied in Sec.~\ref{results} to discuss the implications of the quark mixing renormalization for the determination of the CKM matrix elements $|V_{tq}|$, assuming that they are extracted directly from the top-quark partial decay widths. In particular, we evaluate the relative shifts in the $|V_{tq}|^2$ parameters induced by these effects and discuss their magnitude. Although the $|V_{tq}|$ values obtained in the global analysis of the SM are currently derived taking into account CKM unitarity relations, we employ them as inputs in our calculations with the expectation that they will provide a good approximation to the values determined in the future without invoking such constraints. In Sec.~\ref{ratio}, we apply the results of Sec.~\ref{top-decay} to the evaluation of the ratio ${\cal R}={\Gamma}(t\rightarrow Wb)/\Gamma_t$ and the determination of $|V_{tb}|^2$. We also discuss the scheme dependence in the theoretical calculation of ${\cal R}$. Section~\ref{conclusions} summarizes our conclusions.

\section{Top-quark partial decay widths\label{top-decay}}

In the SM, with three generations of quarks and leptons, the only two-particle decays of the top quark that occur at leading order (LO) are $t\rightarrow Wq$ ($q=d,s,b$). Their partial widths at LO are proportional to the squares of the CKM matrix elements $|V_{tq}|^2$ and are given by
\begin{equation}
\Gamma_0(t\to Wq)=\frac{G_F m_W^2}{4\sqrt{2}\pi}|V_{tq}|^2\frac{\kappa(m_t^2,m_W^2,m_q^2)}{m_t^3}\left[\frac{m_t^2+m_q^2}{2}+\frac{(m_t^2-m_q^2)^2}{2m_W^2}-m_W^2
\right],
\end{equation}
where $G_F$ is the Fermi constant and
\begin{equation}
\kappa(x,y,z)=\sqrt{x^2+y^2+z^2-2(xy+yz+zx)}
\end{equation}
is K\"all\`en's function.

The next-to-leading-order (NLO) QCD~\cite{JK1989,DS1991,EMMS1991} and electroweak corrections~\cite{DS1991,EMMS1991}, were determined quite some time ago, and are included in our calculations.
For completeness, we mention that the corrections due to the finite width of an
off-shell $W$ boson~\cite{JK1993} and the first few terms of the expansion in
powers of $(m_W/m_t)^2$ of the next-to-next-to-leading order (NNLO) QCD corrections~\cite{CM1999,CHSS1999} have also been evaluated, but are not included in this paper. The NLO electroweak corrections are positive, of ${\cal O}(2\%)$, while the NLO and NNLO QCD corrections are negative and of ${\cal O}(10\%)$ and ${\cal O}(2\%)$, respectively.   

In our numerical analysis, we perform the calculations with the aid of the {\tt LOOPTOOLS}~\cite{LT2010} package embedded into the {\tt MATHEMATICA}~\cite{Math7.0} environment. We employ the following input parameters~\cite{PDG2010}:

\vspace*{0.5cm}
\begin{tabular}[b]{l@{\vrule height 12pt depth 4pt width 0pt\hskip\arraycolsep}ll}  
$G_F=1.16637\times 10^{-5}~{\rm GeV}^{-2}$, 
& $\alpha_s^{(5)}(m_Z)=0.1184$, \\  
$m_W=80.399~{\rm GeV}$, & $m_Z=91.1876~{\rm GeV}$, & $m_H=120~{\rm GeV}$, \\  
$m_e=0.510998910~{\rm MeV}$,\ \ & $m_\mu=105.658367~{\rm MeV}$,  
& $m_\tau=1776.82~{\rm MeV}$, \\  
$m_u=2.4~{\rm MeV}$, & $m_d=4.8~{\rm MeV}$, & $m_s=101~{\rm MeV}$, \\  
$m_c=1.27~{\rm GeV}$, & $m_b=4.25~{\rm GeV}$. &   
\end{tabular}  
\vspace*{0.5cm}

\noindent For the top-quark mass, we use the most recent world average
$m_t=173.1~{\rm GeV}$~\cite{:2009ec}, while for $m_u$, $m_d$, and $m_b$, we
employ representative values.
  
We evaluate the CKM matrix elements from the standard parameterization written in terms of $\lambda$, $A$, $\overline{\rho}$, and $\overline{\eta}$~\cite{PDG2010}. This ensures that the CKM matrix is unitary to all orders in $\lambda$. In particular, we employ the values $\lambda=0.2253$, $A=0.808$,  
$\overline{\rho}=0.132$, and $\overline{\eta}=0.341$~\cite{PDG2010}.  

\begin{table}[ht]  
\centering  
{\tiny{  
\begin{tabular}[b]{|l@{\vrule height 12pt depth 4pt width 0pt 
\hskip\arraycolsep}|c|c|c|c|c|c|c|c|c|c|}\hline\hline  
Partial width & Born & Born+QCD &Ref.~\cite{DS1990} & Ref.~\cite{GGM1999} & Ref.~\cite{DK2001} & 
Ref.~\cite{KS2006} & Ref.~\cite{KS2009} & ${\overline {\rm MS}}$-scheme & $\delta V_{ij}=0$ \\ \hline  
$\Gamma(t\to Wd)\times 10^4$ & 1.1133675 & 0.9959853 & 1.0209752 & 1.0209811 & 1.0209291 & 1.0209811 & 1.0209811 & 1.0141341 & 1.0209746 \\  
$\Gamma(t\to Ws)\times 10^3$ & 2.4275072 & 2.1716923 & 2.2280874 & 2.2281001 & 2.2279868 & 2.2281001 & 2.2281001 & 2.2131578 & 2.2280859 \\  
$\Gamma(t\to Wb)$ & 1.4928407 & 1.3354509 & 1.3715786 & 1.3715786 & 1.3715787 
& 1.3715786 & 1.3715786 & 1.3715946 & 1.3715786 \\\hline
$\Gamma_t$ & 1.4953795 & 1.3377222 & 1.3739088 & 1.3739088 & 1.3739088 
& 1.3739088 & 1.3739088 & 1.3739092 & 1.3739088 \\ \hline\hline  
\end{tabular}}}  
\caption{Partial and total decay widths (in GeV) of the top quark evaluated at one loop using the quark mixing renormalization prescriptions of Refs.~\cite{DS1990,GGM1999,DK2001,KS2006,KS2009} and the $\overline{\rm MS}$ scheme. The entries in the last column are obtained by neglecting quark mixing renormalization.\label{tab1}}   
\end{table}  

In Table~\ref{tab1}, we show the one-loop-corrected partial decay widths of the top quark and its total width $\Gamma_t=\sum_{q=d,s,b}\Gamma(t\rightarrow Wq)$ for the selected definitions of the CKM counterterm matrix~\cite{DS1990,GGM1999,DK2001,KS2006,KS2009}. The results for the CKM matrix renormalization conditions proposed in Refs.~\cite{DS1990,GGM1999} were given already in Ref.~\cite{OBSB2001}. We emphasize that we find full agreement with Ref.~\cite{OBSB2001}, provided we adopt the same values for the input parameters. New results are those in the next three columns, which refer to the three genuine on-shell renormalization proposals of Refs.~\cite{DK2001,KS2006,KS2009}, respectively. Note that the prescription of Ref.~\cite{DS1990} leads to a gauge-dependent result, so that the gauge choice must be specified: we perform the calculation in 't Hooft-Feynman gauge, namely the $R_\xi$ gauge with $\xi=1$. For reference, we have included in the previous to last
column of Table~\ref{tab1} the results obtained by renormalizing the CKM matrix in the modified minimal-subtraction ($\overline{\rm MS}$) scheme with 't Hooft mass scale $\mu=m_t$. In order to assess the significance of quark mixing renormalization, we have included in the last column the results of a scheme in which mixing is neglected in loops inserted in the external quark legs, so that the CKM counterterms $\delta V_{ij}$ can be chosen to vanish. It is obtained by replacing $V_{ij}\rightarrow \delta_{ij}$ in the external loops' couplings and replacing the mass of the up-(down-)type virtual quark in such loops by the mass of the external up-(down-)type quark in the $t$-decay process. This diagonal scheme has the virtue of reproducing many of the important contributions of the complete calculations and, at the same time, prevents the emergence of unphysical UV divergences.  

\boldmath
\section{Quark mixing renormalization effects on the determination of $|V_{tq}|$\label{results}}
\unboldmath

At present, the CKM elements $|V_{td}|$ and $|V_{ts}|$ cannot be measured from the respective decays of the top quark, so that one has to rely on determinations from $B^0$--${\overline B}^0$ oscillations mediated by box diagrams involving the top quark, or loop-mediated rare $K$ and $B$ decays. The determination of $|V_{tb}|$ from top-quark decays uses the ratio of partial decay widths ${\cal R}={\Gamma}(t\rightarrow Wb)/\Gamma_t$ (see Sec.~\ref{ratio}). The direct determination of $|V_{tb}|$ without invoking unitarity, but assuming $|V_{td}|,|V_{ts}|\ll|V_{tb}|$, is possible from the single top-quark production cross section. First observations of electroweak single top-quark production were reported by the CDF~\cite{Aal2010} and D0~\cite{Aba2009} Collaborations, leading to:
\begin{equation}
\begin{array}{cccc}
|V_{tb}| & > & 0.71 & [2],\\
|V_{tb}| & > & 0.78 & [5],\\
\end{array}
\end{equation}
at the 95\% confidence level.

\begin{table}[ht]
\centering   
{\footnotesize{   
\begin{tabular}[b]{|c@{\vrule height 12pt depth 4pt width 0pt  
\hskip\arraycolsep}|c|c|c|c|c|c|c||c|}\hline\hline   
$\Delta_{tq}^\alpha$ & Ref.~\cite{DS1990} & Ref.~\cite{GGM1999} & 
Ref.~\cite{DK2001} &  Ref.~\cite{KS2006} & Ref.~\cite{KS2009} & 
${\overline {\rm MS}}$ scheme & $|V_{tq}|^2$~\cite{PDG2010}\\  
\hline   
$td$ & $-6.47\times 10^{-5}$ & $-6.34\times 10^{-4}$ & $4.45\times 10^{-3}$ & $-6.35\times 10^{-4}$ & $-6.34\times 10^{-4}$ & $0.67$ & $7.4304\times 10^{-5}$\\   
$ts$ & $-6.48\times 10^{-5}$ & $-6.34\times 10^{-4}$ & $4.45\times 10^{-3}$ & $-6.34\times 10^{-4}$ & $-6.35\times 10^{-4}$ & $0.67$ & $1.6241\times 10^{-3}$\\   
$tb$ & $1.10\times 10^{-7}$ & $1.08\times 10^{-6}$ & $-7.53\times 10^{-6}$ &  $1.08\times 10^{-6}$ & $1.08\times 10^{-6}$ & $-1.17\times 10^{-3}$ & $0.99831$\\   
\hline\hline   
\end{tabular}}}   
\caption{Relative shifts $\Delta_{tq}^\alpha$ (in \%) in $|V_{tq}|^2$ induced by quark mixing renormalization according to the prescriptions $\alpha$ of Refs.~\cite{DS1990,GGM1999,DK2001,KS2006,KS2009} and the
${\overline {\rm MS}}$ scheme.\label{tab2}}   
\end{table}     

It is important to note that the corrections due to quark mixing renormalization discussed in Sec.~\ref{top-decay} affect also the theoretical calculations of the accurate observables underpinning the determination of the CKM elements $|V_{tq}|$. In order to discuss this matter, we call $\delta_{tq}^\alpha$ the one-loop correction in renormalization scheme $\alpha$ and $\delta_{tq}^0$ the one corresponding to the last column in Table~\ref{tab1}. Taking into account that, in conventional determinations of the $|V_{tq}|$ parameters, the quark mixing effects in the external legs are generally neglected, as is also the case in $\delta_{tq}^0$, we readily find the relation:
\begin{equation}
|V_{tq}^\alpha|^2(1+\delta_{tq}^\alpha)=|V_{tq}|^2(1+\delta_{tq}^0),
\end{equation}
where $\alpha$ labels the renormalization scheme employed. In turn, this implies that
\begin{equation}
\frac{|V_{tq}^\alpha|^2}{|V_{tq}|^2}=R_{tq}^\alpha,
\label{Rtialpha}
\end{equation}
where $R_{tq}^\alpha$ are the ratios of the entries in the last column in Table~\ref{tab1} to those in the $\alpha$ column. Further, Eq.~(\ref{Rtialpha}) permits us to evaluate the relative shifts
\begin{equation}
\Delta_{tq}^\alpha=\frac{|V_{tq}^\alpha|^2-|V_{tq}|^2}{|V_{tq}|^2}=R_{tq}^\alpha-1
\end{equation}
in the $|V_{tq}|^2$ parameters induced by the quark mixing renormalization effects, an issue of considerable interest given the fundamental importance of the CKM parameters. The results are portrayed in Table~\ref{tab2}. (In order to compute the entries in Table~\ref{tab2}, we have used more precise values than displayed in Table~\ref{tab1}.) Note that they are based on the assumption that the CKM matrix elements involving the top quark can be measured directly from top-quark decays. We have seen that, at present, this is not the case for $|V_{td}|$ and $|V_{ts}|$. However, they are of considerable interest for future direct measurements of $|V_{tb}|$ from single top-quark production and, in the case of $|V_{td}|$ and $|V_{ts}|$, if direct determinations from top-quark decays become possible.


From Table~\ref{tab2} we see that, in the renormalization schemes of Refs.~\cite{DS1990,GGM1999,DK2001,KS2006,KS2009}, the relative shifts $\Delta^{\alpha}_{tq}$ induced by the quark mixing renormalization effects are very small, of ${\cal O}(10^{-3}\%)$ or less. In the $\overline{\rm MS}$ scheme, $\Delta^{\overline{\rm MS}}_{tb}$ is still very small, of ${\cal O}(10^{-3}\%)$, while $\Delta^{\overline{\rm MS}}_{td}$ and $\Delta^{\overline{\rm MS}}_{ts}$ are considerably larger, reaching $0.67\%$. This is due to the presence of significant finite corrections that are not removed by the $\overline{\rm MS}$ subtraction.

\boldmath 
\section{The ratio ${\cal R}={\Gamma}(t\rightarrow Wb)/\Gamma_t$ and $|V_{tb}|^2$\label{ratio}}
\unboldmath

$|V_{tb}|$ is the best known CKM matrix element, with relative error 0.0044\%, assuming three generations and employing the unitarity constraints of the CKM matrix. If the assumption of three generations is relaxed, $|V_{tb}|$ is almost completely unconstrained, as~\cite{PDG2010}
\begin{equation}
0.08 \le |V_{tb}| \le 0.9993.
\end{equation}
In the latter scenario, $|V_{tb}|$ was measured at the Tevatron. In particular, the CDF and D0 Collaborations measured at the Tevatron the fraction of bottom-quark events in the total sample of top-quark decays:
\begin{equation}
{\cal R}=\frac{{\Gamma}(t\rightarrow Wb)}{{\Gamma_{t}}}\approx\frac{|V_{tb}|^2}{|V_{td}|^2+|V_{ts}|^2+|V_{tb}|^2}.
\end{equation}
The last term gives the interpretation of this measurement at LO in terms of CKM matrix elements when the down-type quarks are taken to be mass degenerate. If we assume three generations and take into account the unitarity of the CKM matrix, the denominator of this expression is unity, so that ${\cal R}\approx|V_{tb}|^2$. Without this assumption, the measurements of this fraction, which come out close to unity, show that $|V_{td}|,|V_{ts}|\ll|V_{tb}|$, but they do not allow conclusions regarding the absolute magnitude of $|V_{tb}|$. A collection of CDF and D0 results on $\cal R$ is given in Refs.~\cite{Leo2007,Aba2008}, the most recent result being ${\cal R}=0.97_{-0.08}^{+0.09}$~\cite{Aba2008}. 

\begin{table}[ht]
\centering   
{ 
\begin{tabular}[c]{|c@{\vrule height 12pt depth 4pt width 0pt  
\hskip\arraycolsep}|c|}\hline\hline   
  & $\displaystyle{\cal R}$ \\\hline
 Ref.~\cite{DS1990} & 0.99830397\\
 Ref.~\cite{GGM1999} & 0.99830396\\
Ref.~\cite{DK2001} & 0.99830405 \\
Ref.~\cite{KS2006} & 0.99830396\\
Ref.~\cite{KS2009} & 0.99830396\\
${\overline {\rm MS}}$ scheme & 0.99831534\\
$\delta V_{ij}=0$ & 0.99830398\\\hline
Born & 0.99830224\\
\hline\hline  
\end{tabular}
}   
\caption{Ratio ${\cal R}={\Gamma}(t\rightarrow Wb)/\Gamma_t$ evaluated using the quark mixing renormalization prescriptions of Refs.~\cite{DS1990,GGM1999,DK2001,KS2006,KS2009}, the
${\overline {\rm MS}}$ scheme, and the case $\delta V_{ij}=0$. For completeness, the last row shows the result in the Born approximation.\label{tab3}}   
\end{table}     

To assess the significance of quark mixing renormalization, we show in Table~\ref{tab3} the values of ${\cal R}$ obtained from Table~\ref{tab1} using the renormalization prescriptions of Refs.~\cite{DS1990,GGM1999,DK2001,KS2006,KS2009}. For comparison, we include the evaluation in the ${\overline{\rm MS}}$ and $\delta V_{ij}=0$ schemes. Finally, the last row shows the result in the Born approximation. Comparing the different rows, one sees that the scheme dependence among the five prescriptions~\cite{DS1990,GGM1999,DK2001,KS2006,KS2009} is extremely small, of ${\cal O}(10^{-5}\%)$ or less, while the differences with respect to the ${\overline{\rm MS}}$ and $\delta V_{ij}=0$ evaluations are of ${\cal O}(10^{-3}\%)$ and ${\cal O}(10^{-5}\%)$, respectively. The results stress once more the fact that $|V_{tb}|^2$ is not modified to a high degree of accuracy, assuming three generations and invoking the unitarity of the CKM matrix.

\section{Conclusions\label{conclusions}}

In summary, we have studied the numerical effects of several renormalization schemes for the CKM matrix on the partial and total decay widths of the top quark, using the values of the CKM matrix elements obtained in the global analysis of the SM \cite{PDG2010}. We have employed these results to infer the relative shifts in the CKM parameters $|V_{tq}|^2$ due to quark mixing renormalization corrections, assuming that they are measured directly from top-quark decays. Finally, we have discussed the phenomenological implications of these effects on the ratio ${\cal R}={\Gamma}(t\rightarrow Wb)/\Gamma_t$ and thus on the determination of $|V_{tb}|^2$. 

\section*{Acknowledgements}  
The work of A.~A. Almasy and B.A. Kniehl was supported in part by the German Research Foundation (DFG) through the Collaborative Research Centre No.~676 {\it Particles, Strings and the Early Universe---The structure of Matter and Space Time}. The work of A.A. Almasy was also supported in part by the UK Science \& Technology Facilities Council (STFC) under Grant No.\ ST/G00062X/1. The work of B.~A.~Kniehl was also supported in part by German Federal Ministry for Education and Research (BMBF) through Grant No.\ 05H09GUE and by the Helmholtz Association of German Research Centres (HGF) through the Helmholtz Alliance No.~101 {\it Physics at the Terascale}. The work of A. Sirlin was supported in part by the National Science Foundation through Grant No.\ PHY--0758032.  
  


\begin{thebibliography}{99}

\bibitem{Aco2005} D.~Acosta {\it et~al.}\ (CDF Collaboration), Phys.\ Rev.\
  Lett.\ \textbf{95}, 102002 (2005), arXiv:hep-ex/0505091.

\bibitem{Aal2010} T.~Aaltonen {\it et~al.}\ (CDF Collaboration), Phys.\ Rev.\ D \textbf{82}, 112005 (2010), arXiv:1004.1181~[hep-ex].

\bibitem{Aba2006} V.~M.~Abazov {\it et~al.}\ (D0 Collaboration), Phys.\ Lett.\ B \textbf{639}, 616 (2006), arXiv:hep-ex/0603002.
  
\bibitem{Aba2008} V.~M.~Abazov {\it et~al.}\ (D0 Collaboration), Phys.\ Rev.\ Lett.\ \textbf{100}, 192003 (2008), arXiv:0801.1326~[hep-ex].

\bibitem{Aba2009} V.~M.~Abazov {\it et~al.}\ (D0 Collaboration), Phys.\ Rev.\ Lett.\ \textbf{103}, 092001 (2009), arXiv:0903.0850~[hep-ex].

\bibitem{PDG2010} K.~Nakamura {\it et~al.}\ (Particle Data Group), J.\ Phys.\ G \textbf{37}, 075021 (2010).

\bibitem{AFGGHKKLM2007} J. Alwall, R. Frederix, J.-M. G\'erard, A. Giammanco, M. Herquet, S. Kalinin, E. Kou, V. Lemaitre, and F. Maltoni,
Eur.\ Phys.\ J.\ C \textbf{49}, 791 (2007), arXiv:hep-ph/0607115.

\bibitem{AKS2008} A.~A.~Almasy, B.~A.~Kniehl, and A.~Sirlin, Phys.\ Rev.\ D \textbf{79}, 076007 (2009), arXiv:0811.0355~[hep-ph];
\textbf{82}, 059901(E) (2010).

\bibitem{DS1990} A.~Denner and T.~Sack, Z.\ Phys.\ C \textbf{46}, 653 (1990).

\bibitem{GGM1999} P.~Gambino, P.~A.~Grassi, and F.~Madricardo, Phys.\ Lett.\ B \textbf{454}, 98 (1999), arXiv:hep-ph/9811470.

\bibitem{DK2001} K.-P.~O.~Diener and B.~Kniehl, Nucl.\ Phys.\ \textbf{B617}, 291 (2001), arXiv:hep-ph/0109110.

\bibitem{KS2006} B.~A.~Kniehl and A.~Sirlin, Phys.\ Rev.\ Lett.\ \textbf{97}, 221801 (2006), arXiv:hep-ph/0608306;
Phys.\ Rev.\ D \textbf{74}, 116003 (2006), arXiv:hep-th/0612033.

\bibitem{KS2009} B.~A.~Kniehl and A.~Sirlin, Phys.\ Lett.\ B \textbf{673}, 208 (2009), arXiv:0901.0114~[hep-ph].

\bibitem{JK1989} M.~Je\.zabek and J.~H. K\"uhn, Nucl.\ Phys.\ \textbf{B314}, 1 (1989).

\bibitem{DS1991} A.~Denner and T.~Sack, Nucl.\ Phys.\ \textbf{B358}, 46 (1991).

\bibitem{EMMS1991} G.~Eilam, R.~R.~Mendel, R.~Migneron, and A.~Soni, Phys.\ Rev.\ Lett.\ \textbf{66}, 3105 (1991).

\bibitem{JK1993} M.~Je\.zabek and J.~H.~K\"uhn, Phys.\ Rev.\ D \textbf{48}, 1910 (1993), arXiv:hep-ph/9302295.

\bibitem{CM1999} A.~Czarnecki and K.~Melnikov, Nucl.\ Phys.\ \textbf{B544}, 520 (1999), arXiv:hep-ph/9806244.

\bibitem{CHSS1999} K.~G.~Chetyrkin, R.~Harlander, T.~Seidensticker, and
M.~Steinhauser, Phys.\ Rev.\ D \textbf{60}, 114015 (1999), arXiv:hep-ph/9906273.

\bibitem{LT2010} T.~Hahn, \emph{LoopTools User's Guide} (2010),
\url{http://www.feynarts.de/looptools/}.

\bibitem{Math7.0} Wolfram~Research,~Inc., \emph{Mathematica 7.0}, Champaign, IL (2008).

\bibitem{:2009ec}
The Tevatron Electroweak Working Group for the CDF and D0 Collaborations,
\url{http://tevewwg.fnal.gov/},
arXiv:0903.2503 [hep-ex].


\bibitem{OBSB2001} S.~M.~Oliveira, L.~Br\"ucher, R.~Santos, and A.~Barroso, Phys.\ Rev.\ D \textbf{64}, 017301 (2001), arXiv:hep-ph/0011324.

\bibitem{Leo2007} S.~Leone (for the CDF and D0 Collaborations), in Proceedings of \emph{The 15th International Conference on Supersymmetry and the Unification of Fundamental Interactions}, Karlsruhe, Germany, 2007, edited by W.~de~Boer and I.~Gebauer (University of Karlsruhe, Karlsruhe, 2007), p.~76, arXiv:0710.4983~[hep-ex].

\end{thebibliography}
\end{document}